# Coherent control with a short-wavelength Free Electron Laser

K. C. Prince[1,2,3,*], E. Allaria[1], C. Callegari[1], R. Cucini[1], G. De Ninno[1], S. Di Mitri[1], B. Diviacco[1], E. Ferrari[1], P. Finetti[1], D. Gauthier[1], L. Giannessi[1,4], N. Mahne[1], G. Penco[1], O. Plekan[1], L. Raimondi[1], P. Rebernik[1], E. Roussel[1], C. Svetina[1,5], M. Trovò[1], M. Zangrando[1,3], M. Negro[6], P. Carpeggiani[6], M. Reduzzi[6], G. Sansone[6], A. N. Grum-Grzhimailo[7], E.V. Gryzlova[7], S.I. Strakhova[7], K. Bartschat[8], N. Douguet[8], J. Venzke[8], D. Iablonskyi[9], Y. Kumagai[9], T. Takanashi[9], K. Ueda[9], A. Fischer[10], M. Coreno[11], F. Stienkemeier[12], E. Ovcharenko[13], T. Mazza[14], M. Meyer[14],

[1] *Elettra-Sincrotrone Trieste, 34149 Basovizza, Trieste, Italy,* [2] *Molecular Model Discovery Laboratory, Department of Chemistry and Biotechnology, Swinburne University of Technology, Melbourne, 3122, Australia,* [3] *Istituto Officina dei Materiali, Consiglio Nazionale delle Ricerche,34149 Basovizza, Italy,* [4] *ENEA C.R. Frascati, 00044 Frascati, Rome, Italy,* [5] *University of Trieste, Graduate School of Nanotechnology, 34127 Trieste, Italy,* [6] *Dipartimento di Fisica, CNR-IFN, Politecnico di Milano, Piazza Leonardo da Vinci, 32, 20133 Milan, Italy,* [7] *Skobeltsyn Institute of Nuclear Physics, Lomonosov Moscow State University, Moscow, 119991 Russia,* [8] *Department of Physics and Astronomy, Drake University, Des Moines, Iowa 50311, USA,* [9] *Institute of Multidisciplinary Research for Advanced Materials, Tohoku University, Sendai 980-8577, Japan,* [10] *Max Planck Institute for Nuclear Physics, Heidelberg, 69117 Germany,* [11] *ISM, Consiglio Nazionale delle Ricerche, 34149 Basovizza, Italy,* [12] *Physikalisches Institut, Universität Freiburg, 79106 Freiburg, Germany,* [13] *Institut für Optik und Atomare Physik, TU Berlin, Berlin, Germany,* [14] *European XFEL, Albert-Einstein-Ring 19, 22761 Hamburg, Germany.*

**XUV and X-ray Free Electron Lasers (FELs) produce short wavelength pulses with high intensity, ultrashort duration, well-defined polarization and transverse coherence, and have**

been utilised for many experiments previously possible at long wavelengths only: multiphoton ionization[1], pumping an atomic laser[2], and four-wave mixing spectroscopy[3]. However one important optical technique, coherent control, has not yet been demonstrated, because Self-Amplified Spontaneous Emission FELs have limited longitudinal coherence[4-7]. Single-colour pulses from the FERMI seeded FEL are longitudinally coherent[8,9], and two-colour emission is predicted to be coherent. Here we demonstrate the phase correlation of two colours, and manipulate it to control an experiment. Light of wavelengths 63.0 and 31.5 nm ionized neon, and the asymmetry of the photoelectron angular distribution[10,11] was controlled by adjusting the phase, with temporal resolution 3 attoseconds. This opens the door to new short-wavelength coherent control experiments with ultrahigh time resolution and chemical sensitivity.

Coherent control with lasers involves steering a quantum system along two or more pathways to the same final state, and manipulating the phase and wavelength of light to favour this state. This technique represents a major achievement in the quest to understand and control the quantum world. In some cases it is a simple interference effect between transition matrix elements, say $M_1 e^{i\theta_1}$ and $M_2 e^{i\theta_2}$, and can be written as

$$I \sim |M_1 e^{i\theta_1} + M_2 e^{i\theta_2}|^2 = |M_1|^2 + |M_2|^2 + 2|M_1||M_2|\cos(\theta_1 - \theta_2) \qquad (1)$$

where $I$ denotes intensity and $\theta_1 - \theta_2$ the relative phase. The amplitudes of the matrix elements must have similar absolute values to produce significant interference: otherwise the greater of the terms $|M_1|^2$ or $|M_2|^2$ dominates.

To achieve coherent control, longitudinal (i.e. temporal) phase correlation must first be demonstrated and manipulated. There are many ways to do this with optical lasers[12,13], e.g. using bichromatic light, i.e. two overlapping commensurate wavelengths[11,13]. In one implementation, ionization by two first-harmonic photons and one second-harmonic photon is measured, and the anisotropy of the photoelectron angular distribution is observed as a function of the phase difference between the two temporally overlapping wavelengths[14-17]. At optical wavelengths, harmonics are easily generated, e.g. by nonlinear birefringent crystals or third-harmonic generation in gases. In the XUV region, frequency doubling or tripling is impractical due to the lack of efficient media. The phase difference is easily tuned in the optical region by gas cells or mechanical delay lines. In the XUV region, such methods become difficult or impossible, as all gases absorb too strongly to function efficiently, while mechanical delay lines require extreme precision in path length differences. Soft X-ray delay lines are usually constructed with grazing incidence optics, and the required resolution and stability is beyond present technology. In particular, it is very difficult to maintain nanometer and microradian precision in an instrument several meters long. A recent state-of-the-art XUV delay line has a time resolution of 210 attoseconds[18], insufficient for coherent control at short wavelengths, with a much shorter period. Higher performance (40 attoseconds) is possible using normal incidence, split-mirrors[19]. This geometry functions at long wavelengths or over narrow ranges at short wavelengths with special coatings such as multilayers, and filters working in restricted ranges, and so transmission is limited.

High-Harmonic Generation (HHG) sources produce ultrafast pulses of soft X-ray light as a comb of harmonics of the fundamental radiation[20]. Although the coherence of the spectral components of the comb has been verified in several experiments, and is at the basis of the attosecond temporal structure[21], a straightforward and widely applicable method to control the relative phase of two harmonics has not been demonstrated. Also, harmonics generated by HHG do not have the high pulse energy and continuous tunability of FELs. Coherent control using trains of attosecond pulses and synchronized infrared (IR) fields has been demonstrated[22,23] where the control parameter is the relative timing between the attosecond bursts and the phase of the IR field, rather than the relative phase of the XUV harmonics. Bichromatic multi-photon ionization has also been reported[24] with phase control, but again the phase of optical photons was controlled, not that of XUV light.

Here we demonstrate and exploit the longitudinal coherence of two-colour XUV light from FERMI by adopting a radically different approach to tuning the phase: instead of generating the light and manipulating the phase subsequently, two colours are generated by the FEL with a variable phase difference. An *electron* delay line controls the phase of the light, which is adjusted by varying the phase of the electron bunch relative to that of the first colour. The delayed electrons then generate the second colour with a delayed phase. The carrier wave phase and pulse envelope are shifted, but for long pulses (~100 fs), the envelope shift is unimportant.

FERMI has been described[8], and here we summarise the salient points of the machine, Fig. 1. Six APPLE-type undulators[25] can be set independently to produce polarised light at harmonics of the seed wavelength: we used horizontal linear polarisation. Between each pair of undulators, an electron delay line or phase shifter[26] lengthens the path of the electrons by nm scale increments, thus allowing tuning of the relative phase between the bunched electron beam and the co-propagating photon beam (see Methods). This is the key to our approach: $n$ undulators are set to the first harmonic, 6-$n$ are set to the second harmonic, and the phase shifters are used to adjust the phase difference between the harmonics. The temporal and phase profiles were theoretically simulated and both ~100 fs pulses overlap well (see Methods and Supplementary figure 1.)

Figure 2(a) shows the experimental set-up and Figure 2(b) a typical spectrometer image.

The $2s^22p^5(^2P°_{3/2})4s$ resonance of Ne at 62.97 nm (hereafter "4$s$ resonance") was selected and the first five undulators were set to it (see Methods). The sixth undulator was set to radiate at the second harmonic, 31.49 nm, while the electron delay line between the fifth and sixth undulators controlled the relative phase. We checked for spurious effects (see Methods and Supplementary Fig. 2.) The overlapping beams were then transported to the experimental chamber via the PADReS system[27] and focused to a measured spot size of 7-10 μm (see Methods and Supplementary Fig. 3).

The scheme of the experiment is shown in Fig. 3(a). 2$p$ electrons from neon can be emitted by two quantum paths: by a single photon (frequency 2ω) as an *s*- or *d*-wave; or by two photons (frequency ω) as a *p*- or *f*-wave. The weak second-harmonic field ionizes by a first-order process, whereas the intense first-harmonic field ionizes by a second-order process: the ionization rates were adjusted to similar values by varying the intensities of the two wavelengths. Choosing the 4*s* resonance enhances the cross-section for the two-photon process and selects an outgoing *p*-wave, without a significant *f*-wave contribution. Due to the non-linear nature of the process[28] and different parity of the outgoing electronic wave packets generated by the two wavelengths, symmetry breaking occurs in the photoelectron angular distribution with respect to the plane perpendicular to the electric vector of the light, see Fig. 1 of ref. 16. The asymmetry depends strongly on the relative phase of the two fields, and gives rise to an oscillatory term like that in equation (1). If the temporal lag between the two harmonics is $\Delta t$, the relevant parameter is the *delay-induced phase difference* $\Delta\phi=2\omega\Delta t$.

The photoelectron angular distributions were measured using the VMI spectrometer of the Low Density Matter end-station, see Methods. The 'left-right' asymmetry was quantified by the parameter $A_{LR}$,

$$A_{LR} = \frac{I_L - I_R}{I_L + I_R} \qquad (2)$$

where $I_L$ and $I_R$ are the integrated intensities on the left and right of the image.

Figure 3(b) shows the asymmetry parameter $A_{LR}$ as a function of $\Delta\phi$. Clear oscillations are present, with a period $2\pi$ rad or 105 attoseconds, the second harmonic period. The measurement steps were approximately 10 as, but a subsequent scan over a limited range, with steps of 900 zeptoseconds, indicated a resolution of 3.1 attoseconds, Fig. 3(c).

To understand the asymmetry in detail, the angular distribution was fitted with Legendre polynomials[29], each characterised by a β parameter. The even ($\beta_2$, $\beta_4$ etc.) and odd ($\beta_1$, $\beta_3$ etc.) numbered parameters describe the symmetric and antisymmetric parts of the distribution respectively. The fits for $\beta_1$ and $\beta_3$ in Fig. 3(b) are consistent with the calculated errors, while $\beta_2$ shows a larger deviation, possibly indicating systematic errors. Comparing the data qualitatively to the calculated spectra of a simpler system, atomic hydrogen[30], we find the key characteristics are reproduced: $\beta_2$ is constant while $\beta_1$ and $\beta_3$ oscillate, with a phase lag, in this case 1.06 rad. The non-zero lag already follows from lowest-order perturbation theory. For infinite pulses and neglecting non-resonant two-photon transitions (Fig. 3(a)), the lag is derived as $\arg(\frac{2\sqrt{2}}{5} - \frac{D_s}{D_d})$, where $D_s/D_d$ is the ratio of the complex first-order ionization amplitudes into the *s*- and *d*-channels (red arrows in Fig. 3(a)). Frozen-core Hartree-Fock calculations of $D_s$ and $D_d$ predict a lag of approximately 0.55 rad for neon. Thus this theory, whose weak point is the single intermediate state approximation, provides only qualitative predictions for our experimental conditions.

The present result demonstrates phase control at the attosecond level with FERMI, and opens the way for unique experiments in the XUV and soft X-ray region, with complete control of the wavelength, polarization, phase and intensity. FERMI produces light with wavelengths down to 4 nm, providing access to core levels, and thus chemical specificity in coherent control experiments. This is impossible with optical lasers. The extreme time resolution may allow the study of ultrafast phenomena: two colour coherent control experiments have been used to study chemical reactions by manipulating the nuclear wave-packet, but now it is possible to shape the electron wave-packet. The present experiment was performed in the gas phase, but it may be adapted to condensed matter to manipulate electron wave-packet motion in processes important in catalysis, photosynthesis, and solar energy production. As well, the generation of attosecond pulses and pulse trains is based on the coherent control of harmonics, and the way is now open to developing such "pulse sculpting" techniques with FELs, as more than two harmonics can be generated at FERMI by appropriate settings of the undulators. Furthermore, this method may be applicable at SASE FELs if operated in single spike or self-seeding modes, greatly extending the wavelength range beyond that presently available.

**References.**

**Figure legends.**

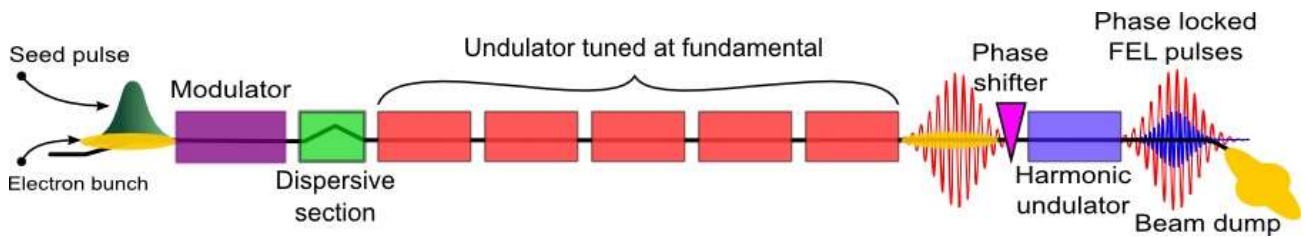

Figure 1. Scheme used in the present study. Red waves indicate schematically the first-harmonic radiation, and blue waves the second-harmonic radiation.

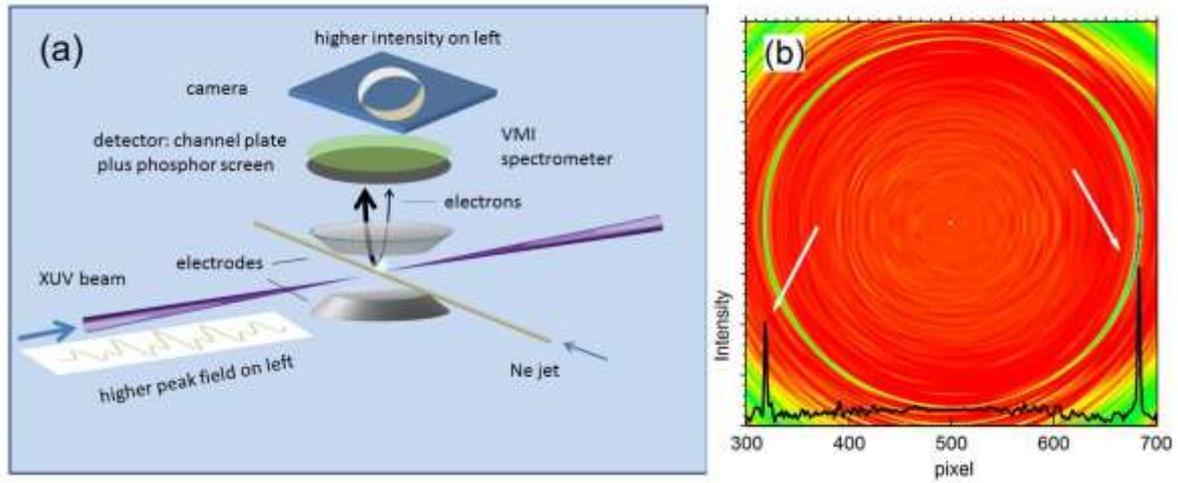

Figure 2. (a) Schematic set-up. The bichromatic light beam with fixed phase relation crosses the atomic jet of neon and ionizes the atoms. The Velocity Map Imaging (VMI) spectrometer measures the angular distribution of ejected electrons. The intensity is higher on the left or right, depending on the phase difference. (b) Typical inverted VMI image, 6000 shots. The strong, sharp ring is due to Ne 2$p$ electrons, emitted by first- and second-harmonic light. A line profile across the centre of the image is shown (black line) at the bottom, demonstrating the left-right asymmetry (white arrows).

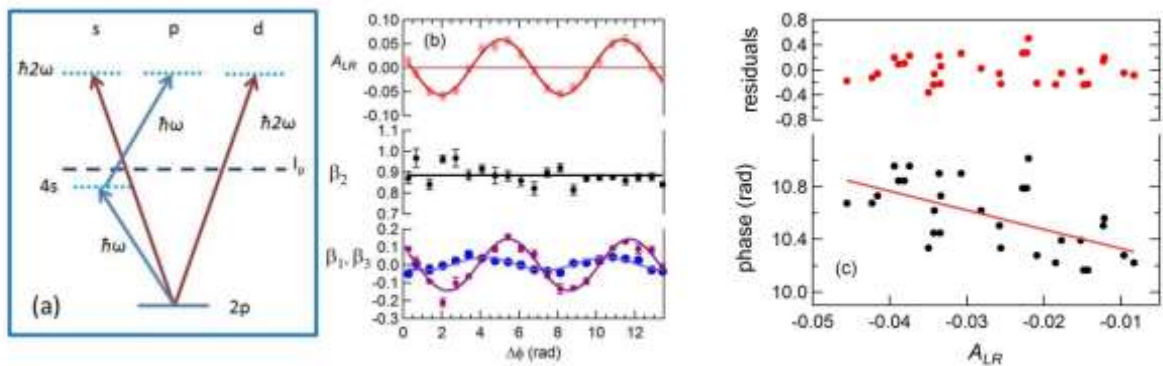

Figure 3. (a) Scheme of the present experiment. A photoelectron may be ejected as a $p$-wave by a two-photon process, or as an ($s+d$)-wave by a one-photon process. (b) Asymmetry parameter $A_{LR}$ as a function of $\Delta\phi$ (red curve), and $\beta_1$ (blue), $\beta_3$ (magenta) and $\beta_2$ (black) parameters as a function of phase. Markers: experimental data; lines: sinusoidal fits for $\beta_1$ and $\beta_3$, linear fit for $\beta_2$. (c) Phase

setting as a function of asymmetry parameter at the steepest part of the delay curve. Step size: 0.056 rad (900 zeptoseconds). Steps were not sequential, so the measurement includes possible errors due to hysteresis in the system, if present. The residuals (difference between the straight line fit and the data) have a standard deviation of the phase corresponding to 3.1 as.

**Author contributions.**

The experiment was conceived by KCP, GS, ANGG and KU, and the method of operating FERMI to carry it out was devised by EA and LG. The experiment was prepared and carried out by KCP, EA, CC, RC, GDN, SDM, BD, EF, PF, DG, LG, NM, GP, OP, LR, PR, ER, CS, MT, MZ, GS, PC, DI, YK, TT, KU, AF, FS, EO, TM, MN, MC and MM. Theoretical calculations (of machine properties or neon spectra) were performed by EA, LG, ANGG, EVG, SIS, KB, ND and JV. Detailed data analysis was performed by MR, PC and DI. The manuscript was drafted by KCP and completed in consultation with all authors.

**Acknowledgements.**

We acknowledge the project CENILS (funded by the Central Europe Programme 2007-2013), which provided the wavefront sensor. PC, MR and GS acknowledge the support of the Alexander von Humboldt Foundation (Project "Tirinto"), the Italian Ministry of Research (Project FIRB No. RBID08CRXK and PRIN 2010ERFKXL_006), and funding from the European Union Horizon 2020 research and innovation programme under the Marie Sklodowska-Curie grant agreement No. 641789 "MEDEA" (Molecular Electron Dynamics investigated by IntensE Fields and Attosecond Pulses). KB, ND and JV acknowledge support from the US National Science Foundation under grants No. PHY-1430245 and XSEDE-090031. DI, YK and KU are grateful for support from the X-ray Free Electron Laser Priority Strategy Program of MEXT. DI, KU and TT are grateful for support from IMRAM, Tohoku University. TM and MM acknowledge support by the Deutsche

Forschungsgemeinschaft (DFG) under grant nos. SFB 925/A1 and A3. ANGG acknowledges support from the European XFEL.

Competing financial interests. The authors declare that they have no competing financial interests.

**Author information**

Reprints and permissions information is available at www.nature.com/reprints. The authors declare no competing financial interests. Correspondence and requests for materials should be addressed to KCP (Prince@Elettra.Eu), GS (giuseppe.sansone@polimi.it) or KU (ueda@tagen.tohoku.ac.jp).

**Methods.**

The measurements were carried out at the Low Density Matter Beamline, FERMI[31, 32]. The relative phase of the two wavelengths was adjusted by means of the electron delay line or phase shifter[26], which is a magnetic structure that deflects the electrons so that they follow a roughly sinusoidal path over two damped periods. The length of the electron trajectory can be increased in small steps from zero to one wavelength of the light (31.49 nm in the present case) or more, with a calibration depending on the electron beam energy and wavelength of the light. In principle the path length difference can be adjusted at the picometer scale, but previous to the present experiment there was no method available for precise determination of the resolution, which we have now measured, see below.

Supplementary figure 1 shows calculated FEL output intensities for the first (63.0 nm, red) and second harmonics (31.5 nm, blue) as a function of time, and their phase difference (black curve), with a quadratic fit (green curve). The simulation was carried out with the FEL time-dependent code Ginger[33], using the undulator and seeding configuration of the experiment shown in Fig. 1 of the article main text. The two pulses have good temporal overlap with calculated pulse durations of 118 and 102 fs. The phase difference has a mild longitudinal dependence on the intensity of the field at the fundamental. The simulation shows that the phase within the FWHM of the pulses has a mean variation of 0.07 rad at 31.5 nm.

The phase shifter introduces a temporal delay between the two pulses, which are otherwise locked in phase by the lasing process. If the temporal lag is $\Delta t$, the phase difference for first plus second harmonic emission is $\Delta \phi = 2*\phi_1 - \phi_2 + 2\omega \Delta t$, where $\phi_1$ and $\phi_2$ are the carrier envelope phases of the two waves. While $\phi_1$ changes randomly for each laser shot, $2*\phi_1 - \phi_2$ is fixed. Consequently, the asymmetry can be controlled by varying $\Delta t$.

The $2s^2 2p^5(^2P°_{3/2})4s$ resonance at 62.97 nm[34] was located by scanning the wavelength of FERMI and measuring the fluorescence yield. The seed laser wavelength was scanned in steps of 50 pm, so that the first harmonic of the FEL wavelength was scanned in steps of 12.5 pm. The central wavelength of the peak in the spectrum was located with an estimated accuracy of 25 pm.

The intensities of the two wavelengths for the experiments were set as follows. With the last undulator open (that is, inactive), the first harmonic from the first five undulators was set to a pulse energy of approximately 50 μJ/pulse, and the two-photon photoelectron signal from neon was observed with the VMI spectrometer. The last undulator was then closed to produce the second harmonic and the photoelectron spectrum of the combined beams was observed: it was much stronger than the two-photon signal, and dominated by the second-harmonic single-photon ionization. The spectrum of the second-harmonic radiation was also monitored using the PADReS spectrometer[35]. Helium was then introduced into the gas attenuation cell of the PADReS system[27], to reduce the intensity of the second harmonic so that the photoelectron intensity was attenuated until a signal was achieved that was equal to about twice that observed by two-photon ionization only with the first harmonic. This corresponded to an attenuation factor of approximately 25. Helium is transparent for the first harmonic. This procedure ensured that the amplitudes of the two ionization pathways were approximately equal.

We checked that the phase shifter did not influence the intensity of the second-harmonic light, and that the second harmonic produced in the first five undulators did not interfere with the measurements using the following method. With the first five undulators tuned to the first harmonic and the sixth tuned to the second harmonic, the spectrum of the second-harmonic radiation was observed by means of the PADReS spectrometer[34] and the phase shifter scanned, see Supplementary Fig. 2. No significant change in the second-harmonic intensity was observed as a function of phase, indicating that negligible phase-coherent second-harmonic light from the first five undulators with linear polarization was present. Different behaviour was observed for a first-plus third-harmonic configuration (data not shown), where the third harmonic was produced coherently on axis in the first undulators: strong interference effects between the light from the earlier and later undulators were observed when they were tuned to the third harmonic, thereby demonstrating that this diagnostic is effective. The use of a helical undulator configuration has been shown to remove this contamination, allowing the flexibility of phase control of odd and even harmonics.

The optical focusing conditions were simulated for optimal curvature of the Kirkpatrick-Baez active optics[27], and verified experimentally by using a Hartmann wavefront sensor. Supplementary Fig. 3 shows the calculated focal spot shape and line profiles of the first and second harmonics, both calculated and measured from the reconstructed image measured by the wavefront sensor. The measured spot size was 7-10 μm (FWHM). The average pulse energy of the first harmonic was 50 μJ, and after correcting for the transmission efficiency of the beam transport optics, we estimate that the average intensity in the focal spot was $1 \times 10^{15}$ W/cm$^2$, for a pulse duration of 100 fs[36].

The atomic beam was produced by a supersonic expansion and defined by a skimmer and vertical slits. The length of the beam along the light propagation direction was approximately 1 mm. The data from the wavefront sensor was used to reconstruct the spot size at the centre and extremes of this excitation volume and compared to spot profiles calculated using the WISE code[37], and other software. The diameter at the extremes is approximately 2% larger than at the centre.

The data was analysed by firstly correcting the VMI images for spatial variations of sensitivity of the detector. The spectrometer works by projecting the expanding sphere of photoelectrons of a given kinetic energy onto a plane surface, and to recover the original angular distribution, the projected images must be inverted. Cylindrical symmetry is assumed and we used the pBASEX software[29]. The inverted (or reconstructed) angular distributions were integrated over the left and right sides of the image to give a value of the asymmetry parameter. An example of an image and a line profile is shown in Fig. 2(b).

The error bars of $A_{LR}$ in Fig. 3(c) of the main text were estimated from the standard deviation of the fitting parameters for the photolines separately for the left and right sides of the VMI images. The error bars for the $\beta_n$ parameters were determined by dividing the data into five subsets, and analysing them separately. Then the standard deviation was calculated for the whole data set. This procedure estimates statistical fluctuations, but does not take account of systematic errors. The $\beta_2$ curve was fitted by a constant straight line and $\beta_1$ and $\beta_3$ by sinusoidal functions, with the frequency of both curves constrained to be equal. A constant for $\beta_2$ and a sine function for odd $\beta$ values are the forms expected for the case of hydrogen[30].

**References**.

**Supplementary Figures and captions**

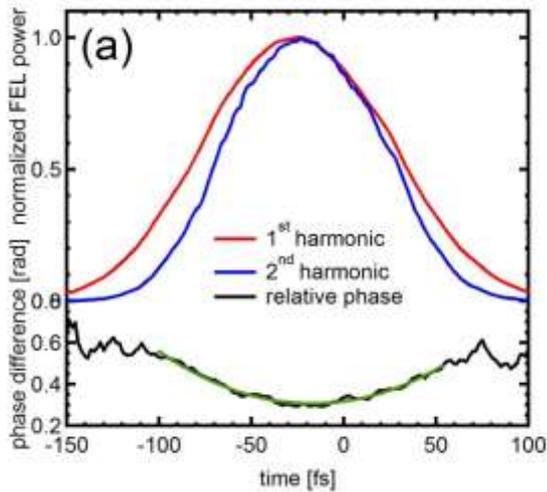

Supplementary figure 1. (a) Calculated intensities of the first and second harmonics (red and blue curves) as a function of time, and their phase difference (black curve), with a quadratic fit (green curve).

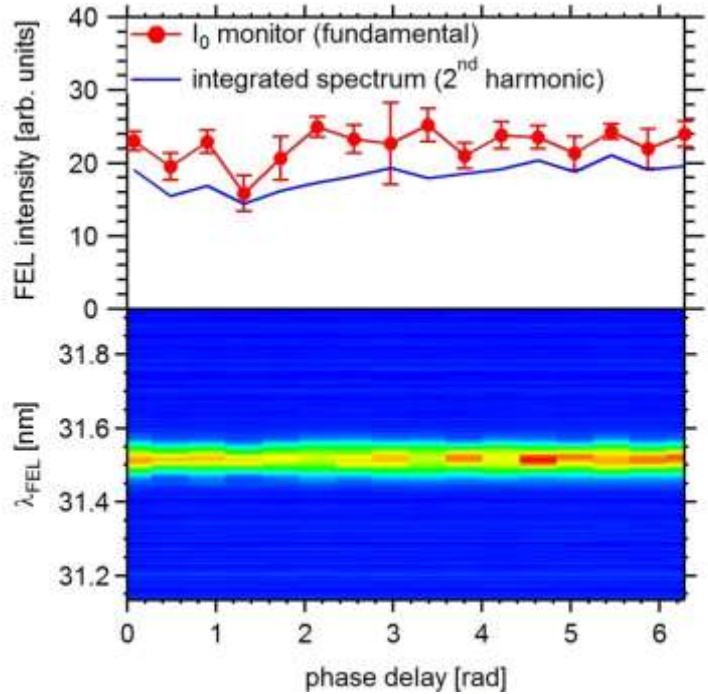

Supplementary figure 2. Lower panel: spectrum averaged over 21 shots is shown on the vertical axis, against phase delay (the zero is arbitrary), in radians. Colour scale indicates intensity (red is high, blue is low). Upper panel: intensity of the first harmonic (I0 monitor, red), monitored by the gas absorption intensity monitor of PADReS, and intensity of the second harmonic, determined from the PADReS spectrometer (blue curve). Error bars indicate the rms variation for each measured point.

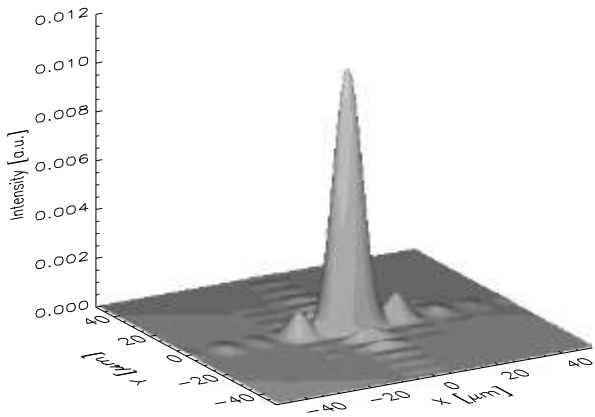

(a)

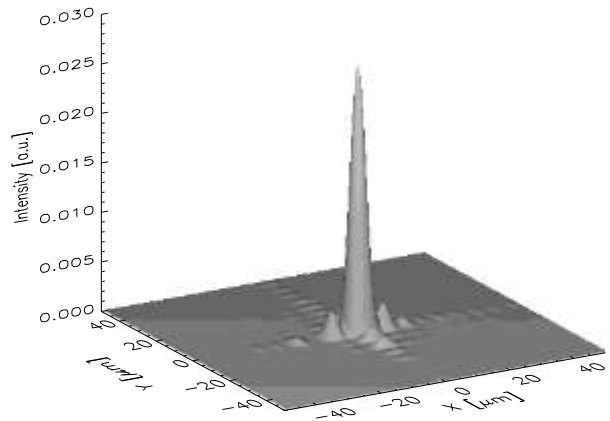

(b)

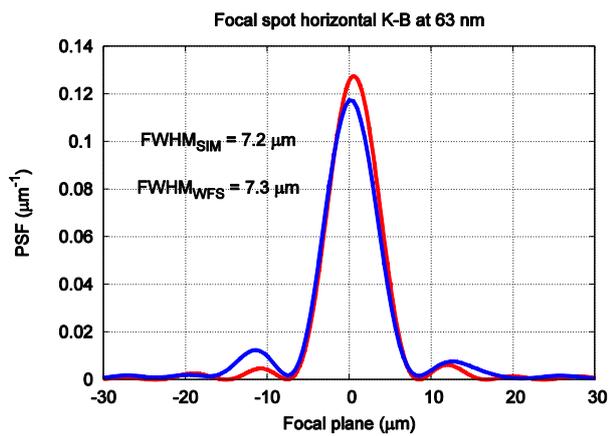

(c)

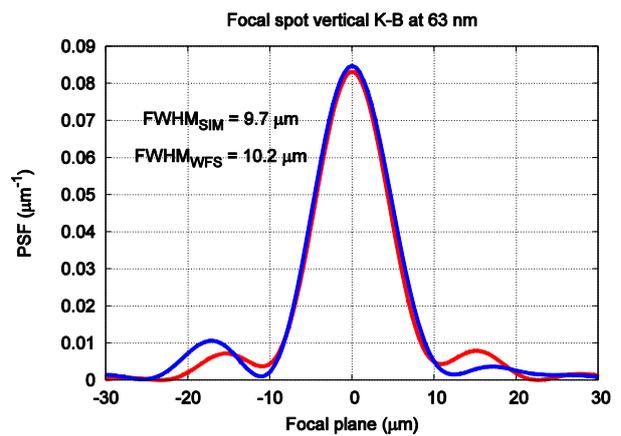

(d)

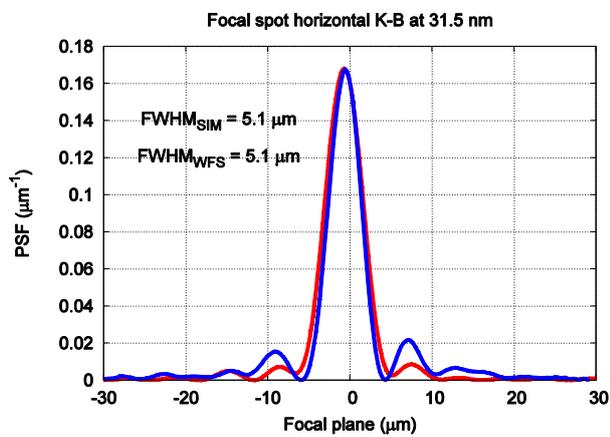

(e)

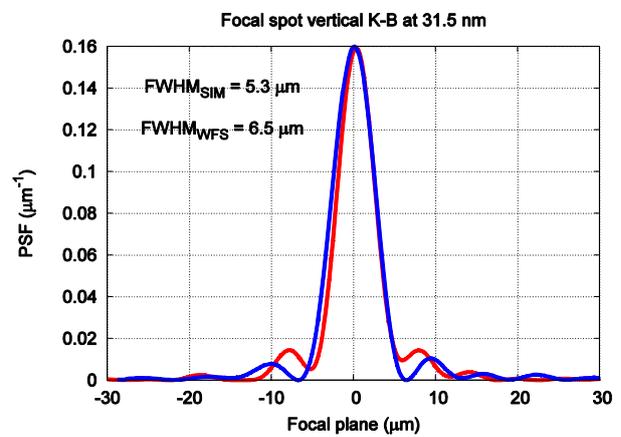

(f)

Supplementary figure 3. Three-dimensional rendering of the calculated focus of the (a) first- and (b) second- harmonic spot profiles. The weak extra peaks are due to diffraction effects, caused by the finite size of the optical elements and apertures. Calculated and reconstructed (from the wavefront sensor) line profiles of the intensity of the first harmonic in the horizontal (c) and vertical (d) directions. The reconstructed profiles are slightly broader than the theoretical values, with a little more diffraction. Calculated and reconstructed line profiles of the intensity of the second harmonic in the horizontal (e) and vertical (f) directions. Good agreement with the theoretical profiles is observed. The two beams overlap well in terms of both position and width of the focal spots. The first harmonic is slightly broader than the second harmonic. The FWHM are given for the measurement (WFS) and simulation (SIM).